\documentclass[showpacs,preprintnumbers,amsmath,amssymb,superscriptaddress,groupedaddress]{revtex4-1}

\usepackage{graphics}% Include figure files
\usepackage{epsfig}

\begin{document}

\title{Mobility of nonlocal solitons in fading optical lattices}

\author{Zhiping Dai}\email{daizhi169@yahoo.com.cn}
\author{Xiaohui Ling}
\author{Youwen Wang}
\author{Kaiming You}
\affiliation{Department of Physics and Electronic Information
Science, Hengyang Normal University,
 Hengyang 421008, P.R. China}

\begin{abstract}
We study the soliton mobility in nonlocal nonlinear media
 with an imprinted fading optical lattice.
The results show that the transverse mobility of solitons varies
with the lattice decay rate and the nonlocality degree,
which provides an opportunity for all-optical control of light.
\end{abstract}


\maketitle

\section{Introduction}\label{introduction}
In these years, the propagation of solitons
in optical lattices has attracted a lot of attention for
its unique properties for
all-optical control of light~\cite{Kartashov-po-2009}.
Previous studies include diffraction-managed solitons in curved
 lattices~\cite{Szameit-pra-2008},
 soliton switching in quasi-continuous
optical arrays~\cite{Kartashov-ol-2004}, and soliton control in chirped optical
lattices~\cite{Kartashov-josab-2005}. Longitudinal modulation of lattices offers additional opportunities
for soliton control. In particular, periodic longitudinal
modulation of the coupling strength results
in soliton oscillations and decay~\cite{Pertsch-Chaos-2003}, harmonic longitudinal
modulation of the linear refractive index leads to parametric amplification of transverse
soliton swinging~\cite{Kartashov-ol-2004-2}, while longitudinal modulation of the nonlinearity can be used
to steer solitons~\cite{Assanto-prl-2010}. Nowadays the technique in fabricating
a desired optical lattice is mature. Even the three-dimensional optical lattice has been created
with the optical induction technique~\cite{Zhang-oe-2009,Xavier-ol-2009}.

Properties of solitons supported by media with local nonlinear response are
now well established. However, under appropriate conditions, the nonlinear
response of materials can be highly nonlocal, a phenomenon
that importantly affects the properties of solitons
supported by such media~\cite{Krolikowski-Quantum-2004}. Some unique properties
are discovered in this context. For example, the nonlocality of the nonlinear response can
affect the soliton mobility profoundly~\cite{Xu-prl-2005,Dai-pra-2008}. Recently, the defect solitons in
parity-time symmetric potentials with nonlocal nonlinearity are investigated.
 It is found that nonlocality can expand the stability region of
 defect solitons~\cite{Hu-pra-2012}.

\section{soliton mobility}\label{model}
In this paper, we study the soliton mobility
in nonlocal nonlinear media with an optical lattice that fades away
exponentially along the propagation direction.
We consider the following dimensionless  nonlocal nonlinear
Schr$\ddot{o}$dinger equation (NNLSE)~\cite{ref15,Guo-pre-2004}
with a fading
 optical lattice~\cite{Dai-cpb-2013}
\begin{equation}\label{NNLSE}
  i\frac{\partial \widetilde{U}}{\partial z}+\frac{1}{2}\frac{\partial^2 \widetilde{U}}{\partial^2
  x}
 +\widetilde{U}\int R(x-\xi){|\widetilde{U}(\xi,z)|}^2
 d\xi+\widetilde{U} p Q(x,z)=0,
\end{equation}
where $\widetilde{U}(x,z)$ is the complex amplitude envelop of the light beam,
$x$ and $z$ are transverse and longitudinal coordinates scaled to
the beam width and diffraction length, respectively, $R(x)$ is the
real symmetric nonlocal response function, and
$pQ(x,z)=p\cos^{2}\big(\Omega x\big)\exp(-\delta z)$ describes the
profile of the optical lattice ($p$ is the lattice depth, $\Omega$
is the lattice frequency,
and $\delta$ is the decay rate) (Fig. 1). In practice, such a kind
of lattice can be induced optically. Specifically, the refractive
index modulation in the transverse direction can be induced with two
interfering plane waves~\cite{Fleischer-prl90-2003,Yang-OE-2008}.
Longitudinal modulation can be created by varying
intensities~\cite{Zhang-OL-2010}, intersection angles, or carrying
wavelength~\cite{Bergstrom-jas-2002} of the lattice-forming plane waves.

As previously indicated in Ref.~\cite{Guo-pre-2004}, in the
strongly nonlocal case if $R(x)$ is twice differentiable at $x=0$,
 Eq.~(\ref{NNLSE}) can be reduced to the following equation
\begin{eqnarray}\label{NNLSE-22}
i\frac{\partial \tilde{U}}{\partial z}+\frac{1}{2}\frac{\partial^2 \tilde{U}}{\partial^2 x}
 +\tilde{U}\int [R_0+\frac{1}{2}R''_0(x-\xi)^2]{|\tilde{U}(\xi,z)|}^2
 d\xi
 +\tilde{U}p \cos^{2}\big(\Omega x\big)\exp(-\delta z)=0,
\end{eqnarray}
where $ R_0=R(0) $ and $ R''_0=R''(0) $.
For the NNLSE, the beam power is an invariant~{\cite{Yakimenko-pre-2006}}
\begin{equation}\label{Power}
 P=\int{|\tilde{U}|}^2 dx.
\end{equation}
If we define the beam center as~\cite{Ouyang-pra-2007}
\begin{equation}\label{Center}
 q(z)=\frac{\int\xi{|\tilde{U}(\xi,z)|}^2 d\xi}{\int{|\tilde{U}(\xi,z)|}^2
 d\xi},
\end{equation}
by making use of Eqs.~(\ref{Power}) and
(\ref{Center}), Eq.~(\ref{NNLSE-22}) is reduced to
\begin{eqnarray}\label{NNLSE-3}
 \nonumber i\frac{\partial \tilde{U}}{\partial z}+\frac{1}{2}\frac{\partial^2 \tilde{U}}{\partial^2 x}
 +\tilde{U}R_0P+\frac{1}{2}\tilde{U}R''_0P{(x-q)}^2+\frac{1}{2}\tilde{U}R''_0\int{(\xi-q)}^2d\xi
 \\
 +\tilde{U}p \cos^{2}\big(\Omega x\big)\exp(-\delta z)=0.
\end{eqnarray}
By the transformation~\cite{Ouyang-pra-2007}
\begin{equation}\label{Transformation}
U=\widetilde{U}\exp\{-i[R_{0}P+\frac{R_{0}^{''}}{2}\int_{0}^{z}dz^{'}\int(\xi-q)^{2}|\tilde{U}(\xi,z)|^{2}d\xi]z\},
\end{equation}
Eq.~(\ref{NNLSE-3}) turns into
\begin{equation}\label{NNLSE-4}
i\frac{\partial U}{\partial
z}+\frac{1}{2}\frac{\partial^{2}U}{\partial x^{2}}-\frac{U}{2}\gamma
P(x-q)^{2}+Up\cos^{2}\big(\Omega x\big)\exp(-\delta z)=0,
\end{equation}
where $\gamma=-R''_0>0$.
\begin{figure}
 \centering
  \includegraphics[width=6.9cm]{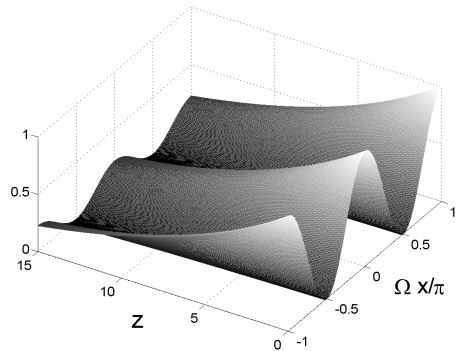}
\\
\caption{Profile of a fading optical lattice with $\delta$ =0.1 and
$p$=1.}\label{fig1}
\end{figure}

Comparing Eq.~(\ref{NNLSE-4}) with the one-dimensional Schr$\ddot{o}$dinger equation
~{\cite{Greiner-Quantum Mechanics-2001}}
\begin{equation}\label{Schrodinger}
i\hbar\frac{\partial }{\partial
t}\psi(x,t)=\big[-\frac{\hbar^{2}}{2m}\frac{\partial^{2}}{\partial
x^{2}}+V(x)\big]\psi(x,t),
\end{equation}
in an equivalent quantum system with $\hbar=1$ and $m=1$, the
equivalent potential energy of Eq.~(\ref{NNLSE-4}) is
\begin{equation}\label{Potential}
V(x)=\frac{1}{2}\gamma P(x-q)^{2} -p\cos^{2}\big(\Omega x\big)\exp(-\delta z).
\end{equation}
With the aid of Ehrenfest theorem~\cite{Greiner-Quantum Mechanics-2001}
and Eqs.~(\ref{NNLSE-4})-- (\ref{Potential}), we get an equation
   of motion for the beam center as follows
\begin{equation}\label{Center equation}
\quad\frac{d^{2}q}{dz^{2}}=\frac{\int{[-\gamma P(x-q)-\Omega p \exp(-\delta z)\sin(2\Omega x)]|U|^{2}}dx}{\int{|U|^{2}}dx}.
\end{equation}
In the context of strongly nonlocal cases, we employ the
following Gaussian-type trial solution
\begin{equation}\label{trial solution}
 U(x,z)=A
\exp[-\frac{{(x-q)}^2}{2w^2}]\exp[i\beta(x-q)+i\phi],
\end{equation}
where $A$ and $\phi$ are the amplitude and phase of the beam respectively,
  $w$ is the beam width, and
$\beta$ is
the incident angle with respect to $z$-axis. This choice of ansatz is justified
since, in the limit $p\rightarrow0$, Eq.~(\ref{trial solution})
describes solitons in uniform strongly nonlocal nonlinear
media~\cite{Guo-pre-2004}.
Substitution of Eq.~(\ref{trial solution}) into Eq.~(\ref{Center
equation}) yields
\begin{equation}\label{Center equation-2}
\quad\frac{d^{2}q}{dz^{2}}+\frac{\Omega_{0}^{2}}{2\Omega}\sin(2\Omega
q)=0,
\end{equation}
where $\Omega_{0}=[2p\Omega^{2}\exp(-\delta
z)\exp(-\Omega^{2}w^{2})]^{1/2}$ defines the frequency of
small-amplitude oscillations of the soliton.

Without loss of generality, we
assume that a soliton is launched into
the medium at the point $x=0$ with an angle $\beta_{0}$, so that
$q|_{z=0}=0$ and $dq/dz|_{z=0}=\beta_{0}$.
For $\delta=0$, it has been shown
 that the soliton's movement is separated by
a critical angle
$\beta_{c}=\big[2p\exp(-\Omega^{2}w^{2})\big]^{1/2}$. When the
incident angle is smaller than the critical angle,
the soliton is trapped within the waveguide
induced by the optical lattices. Above the
critical angle, the soliton leaves the waveguide and
 travels across the lattices~\cite{Dai-pra-2008}.

In the presence of a longitudinal  modulation, i.e., $\delta\neq0$,
 the critical angle
$\beta_{c}\sim\Omega_{0}/\Omega=\big[2p\exp(-\delta
z)\exp(-\Omega^{2}w^{2})\big]^{1/2}$, indicating that $\beta_{c}$
decreases with the propagation distance. As a result, even if the
soliton is launched into the lattice with an incident angle
$\beta_{0}<\beta_{c}$, it will escape from the central waveguide
once the condition $\beta_{0}>\beta_{c}$ is
satisfied.  In the limit $\delta\rightarrow0$, the soliton
is trapped by the first waveguide and never escapes from the central
waveguide. On the contrary, it propagates along its original
direction traveling across the lattices without oscillations.
 As the critical angle depends on the lattice decay rate, a
modulation of the  decay rate leads to a change of the soliton
transverse mobility, a property that might be used  for soliton
steering.

\section{Numerical Simulation}\label{Simulation}
 \begin{figure}
\centering
\includegraphics[width=4.9cm]{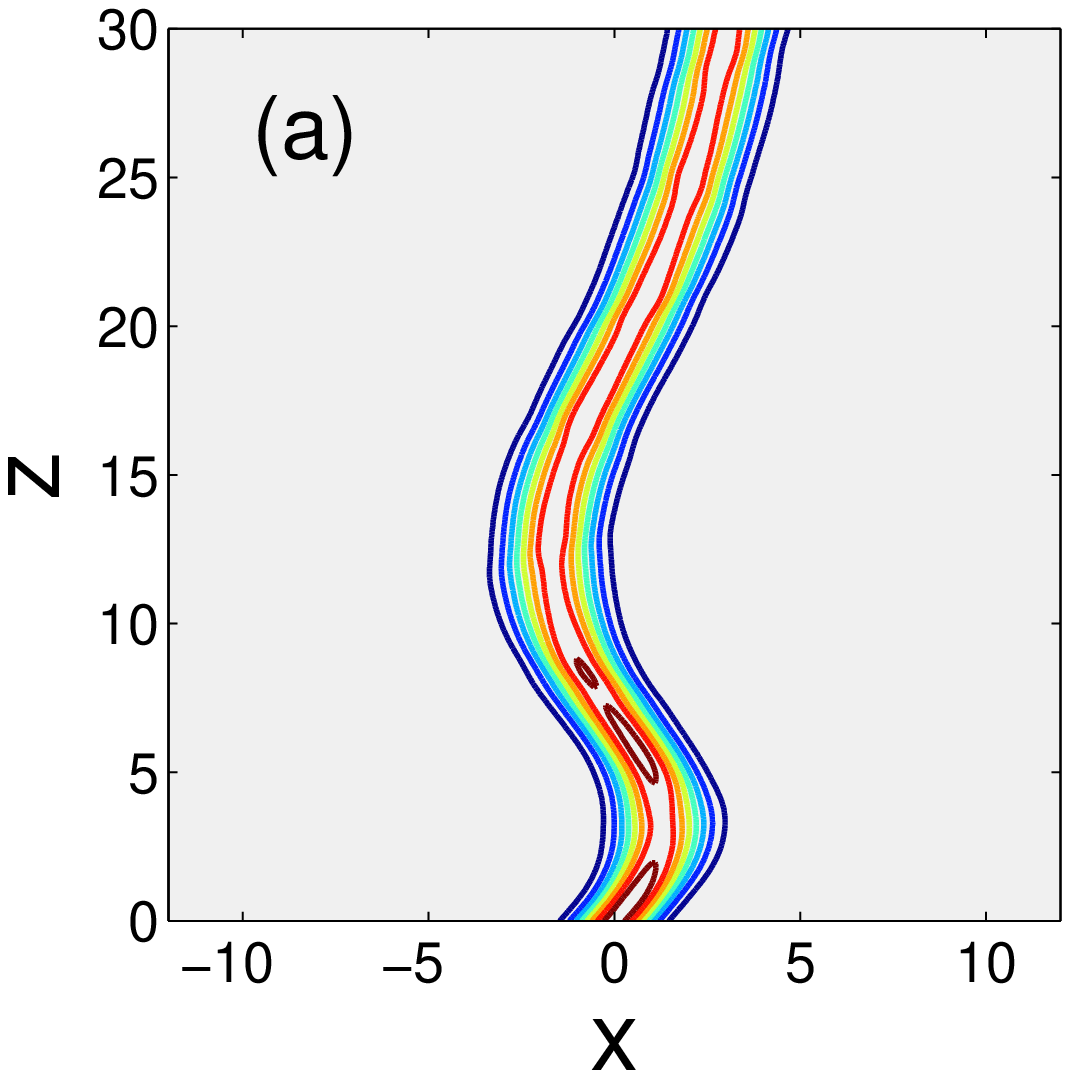}%
  \includegraphics[width=4.9cm]{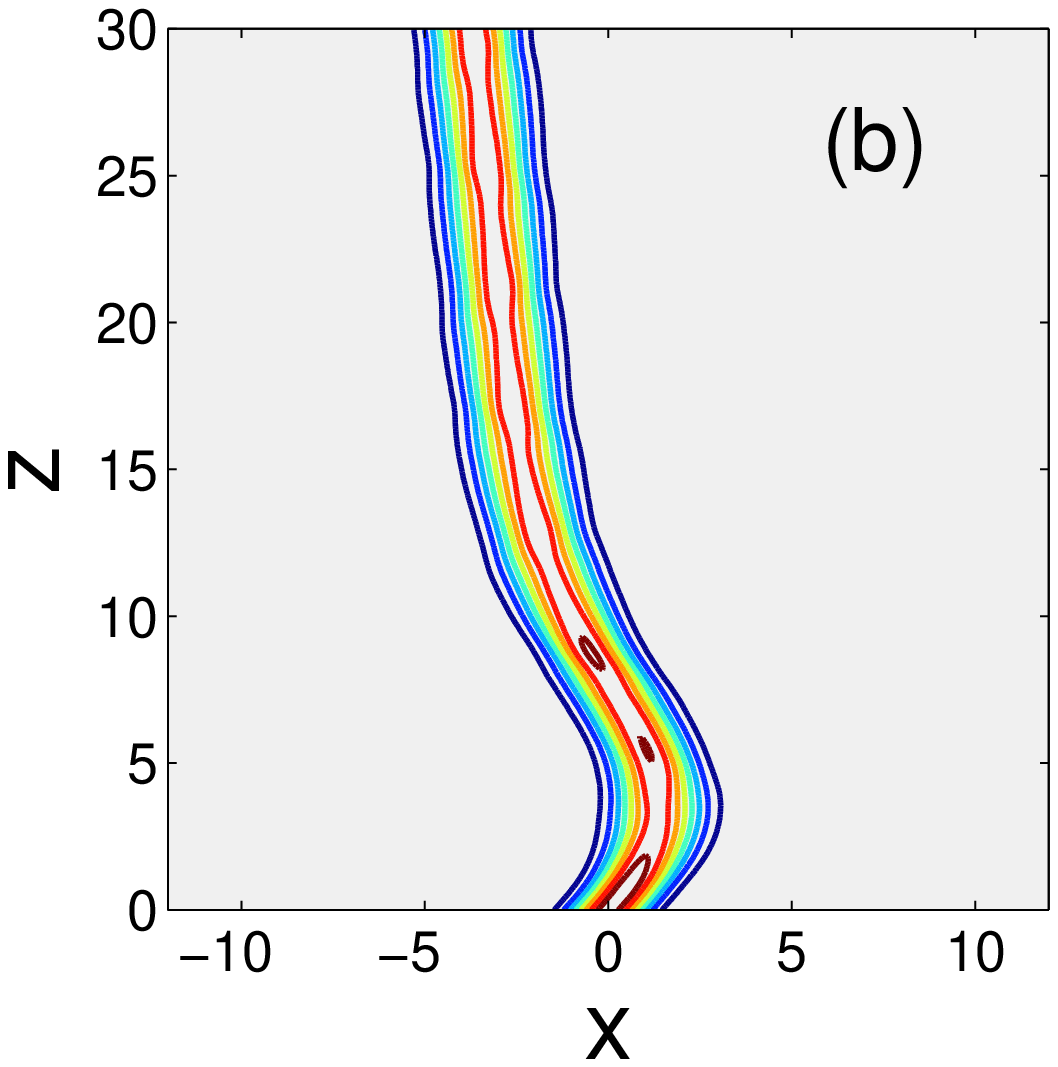}\\
   \includegraphics[width=4.9cm]{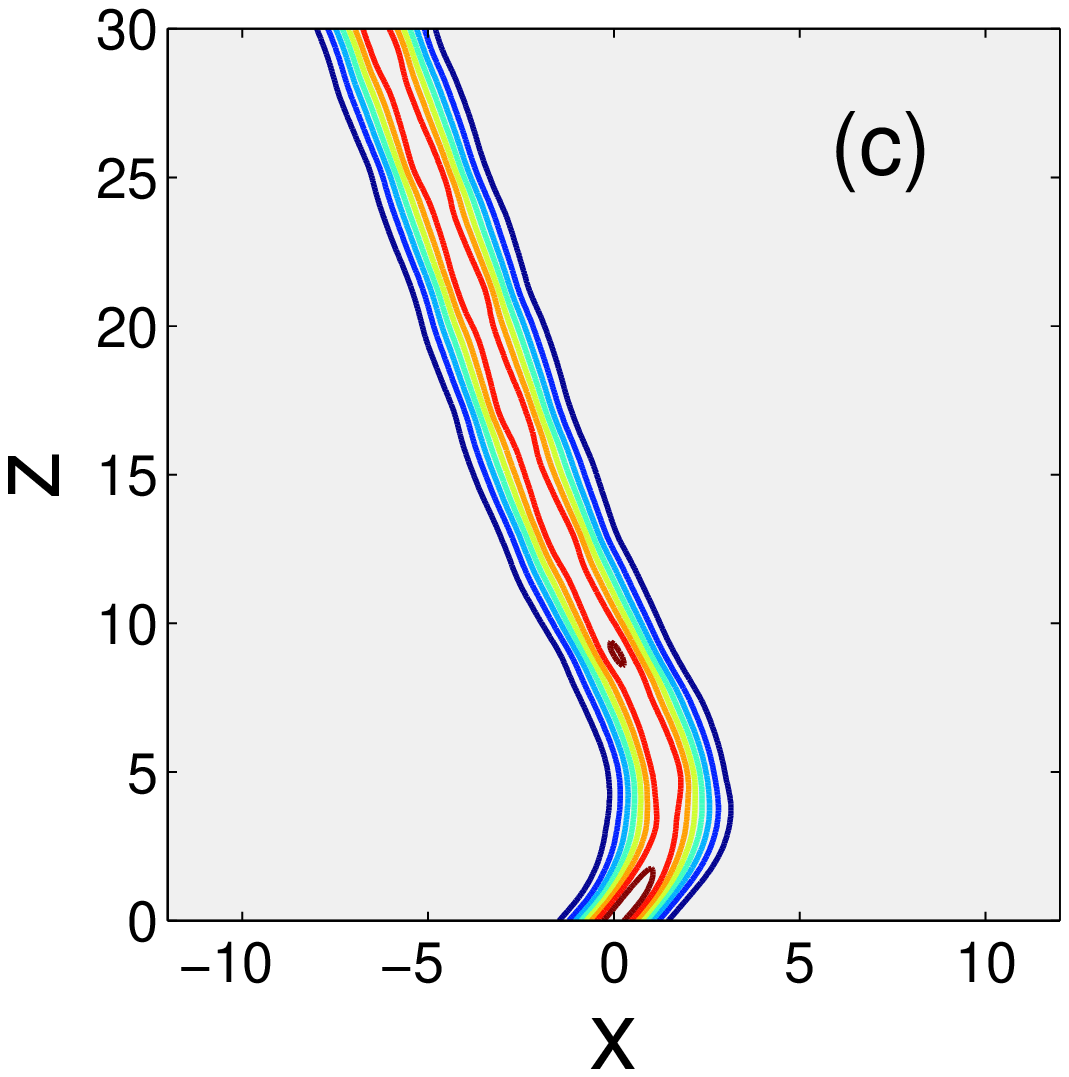}%
  \includegraphics[width=4.9cm]{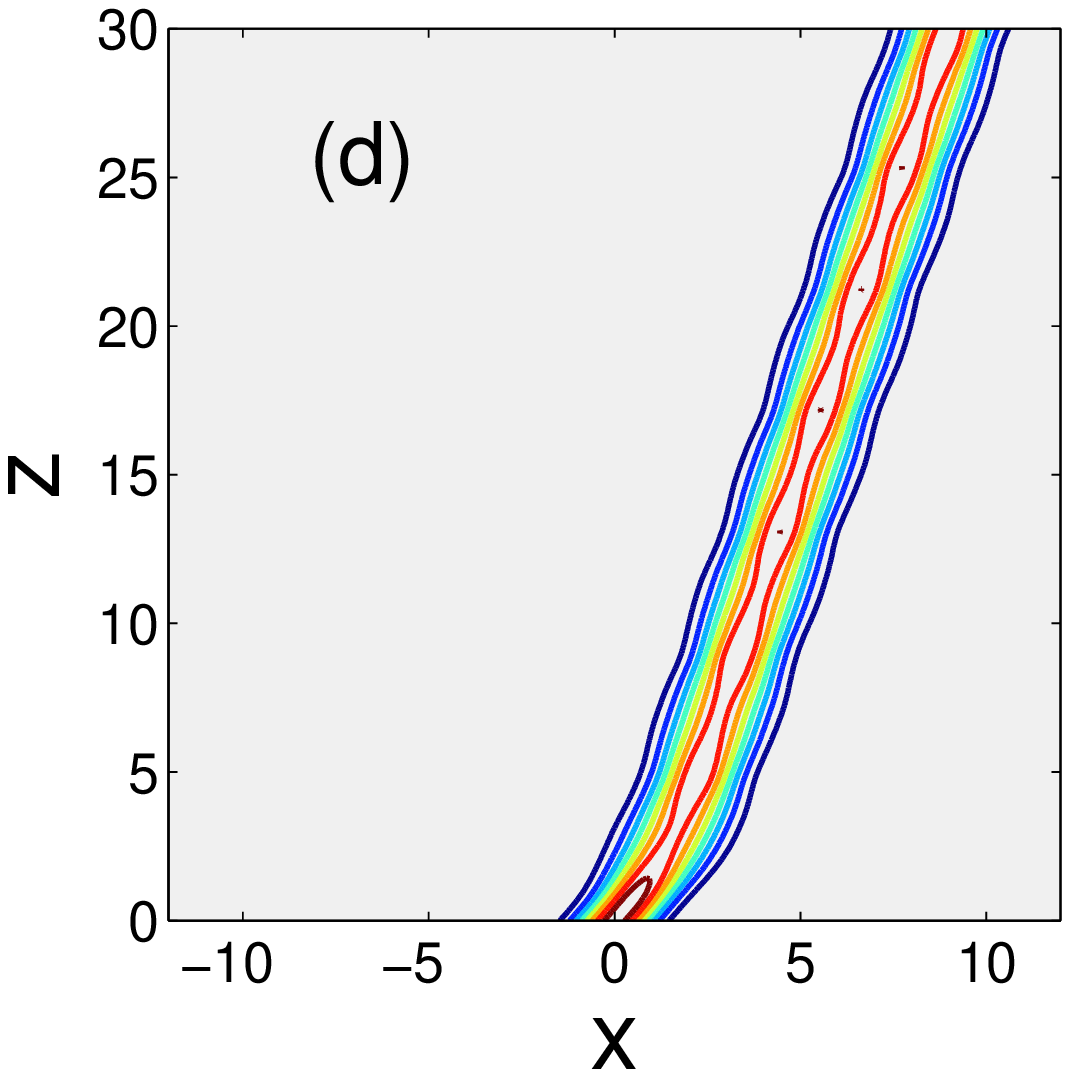}\\
  \caption{Contour plots showing the dependence of the soliton
 mobility on the
lattice decay rate. (a) $\delta$=0.1; (b) $\delta$=0.15;
 (c) $\delta$=0.2;  (d) $\delta$=0.5.
     For all the cases, $\alpha=0.1$, $p$=1, $\Omega$=0.5, and $\beta_{0}$=0.7.}\label{fig3}
\end{figure}

  To intuitively show the effect of the lattice and the nonlocality, we numerically simulate the
propagation of a Gaussian-type beam in nonlocal nonlinear
media with a fading optical lattice. We take
 Eq.~(\ref{NNLSE}) as the evolution equation and $U(x,z)|_{z=0}$ in
 Eq.~(\ref{trial solution})  as the initial condition. The numerical
 arithmetic we use here is the split-step Fourier
 Method~\cite{Agrawal-Nonlinear Fiber Optics-2007}.
 The nonlocal response function is supposed to be a Gaussian function~\cite{Guo-pre-2004},
 i.e.,
$R(x)=\exp(-x^2/2w_m^{2})/(\sqrt{2\pi}w_m)$, where we introduce a nonlocal
parameter $\alpha$ to define the degree of nonlocality
($\alpha$ is the ratio of the beam width $w$ to the characteristic
nonlocal response length $w_m$ of the medium). The smaller the
nonlocal parameter, the stronger the nonlocality degree.

The dependence of the soliton mobility on the lattice decay rate when $\alpha=0.1$ is shown in Fig. 2,
which indicates that the soliton is emitted at different transverse
positions with the change of the lattice decay
rate. Since the critical angle decreases more quickly
 for a larger decay rate, an increase of the rate leads to a decrease
of the number of soliton oscillations, just as predicted in Sec. 2. Depending on
the lattice decay rate, solitons with the angle
$\beta_{0}<\beta_{c}$ perform
different numbers of oscillations and then start walking freely in
diverse directions. When the rate increases to 0.5, the
soliton does not oscillate but is always deflected in the same
direction (Fig. 2(d)).
\begin{figure}
\centering
\includegraphics[width=4.7cm]{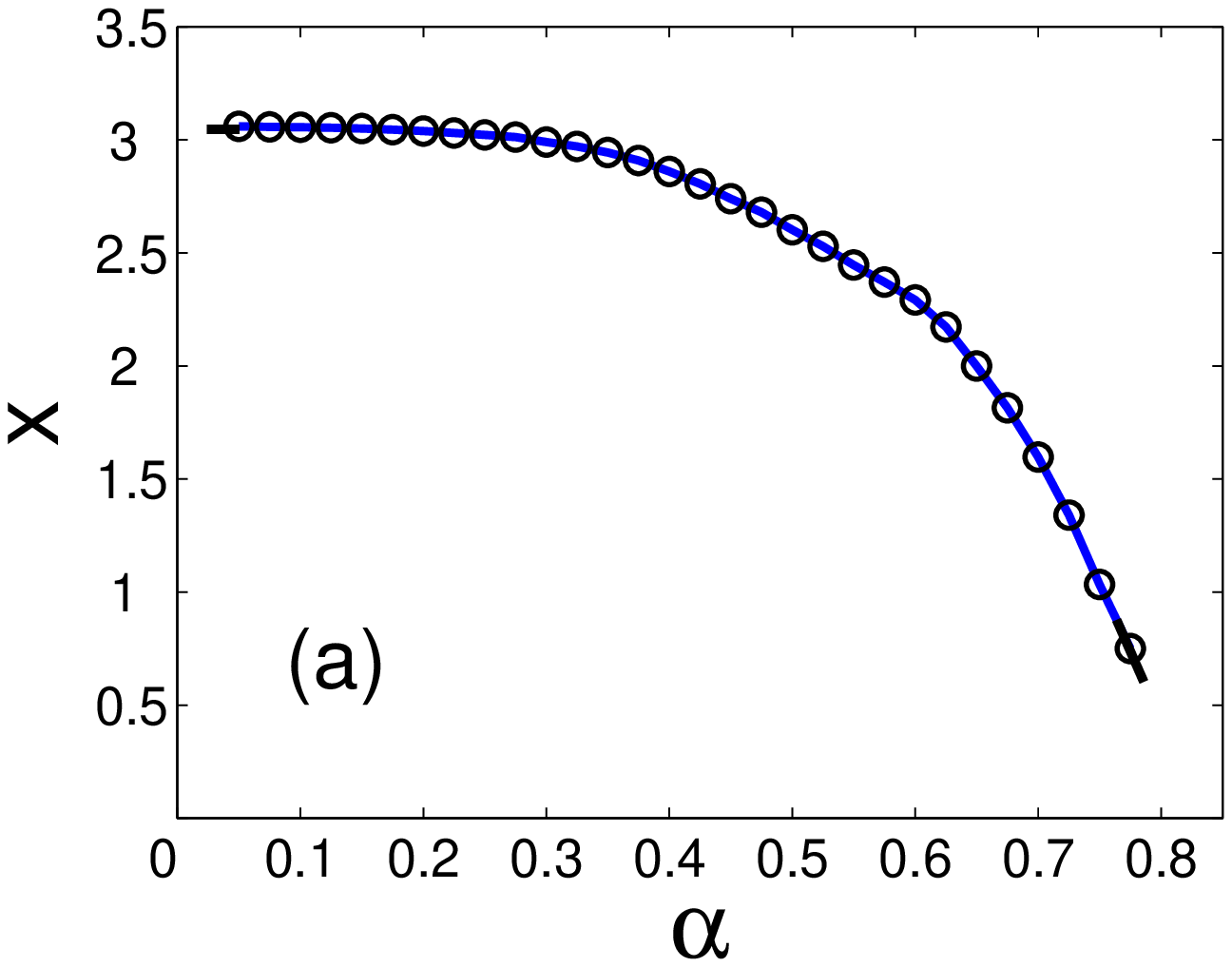}%
  \includegraphics[width=4.7cm]{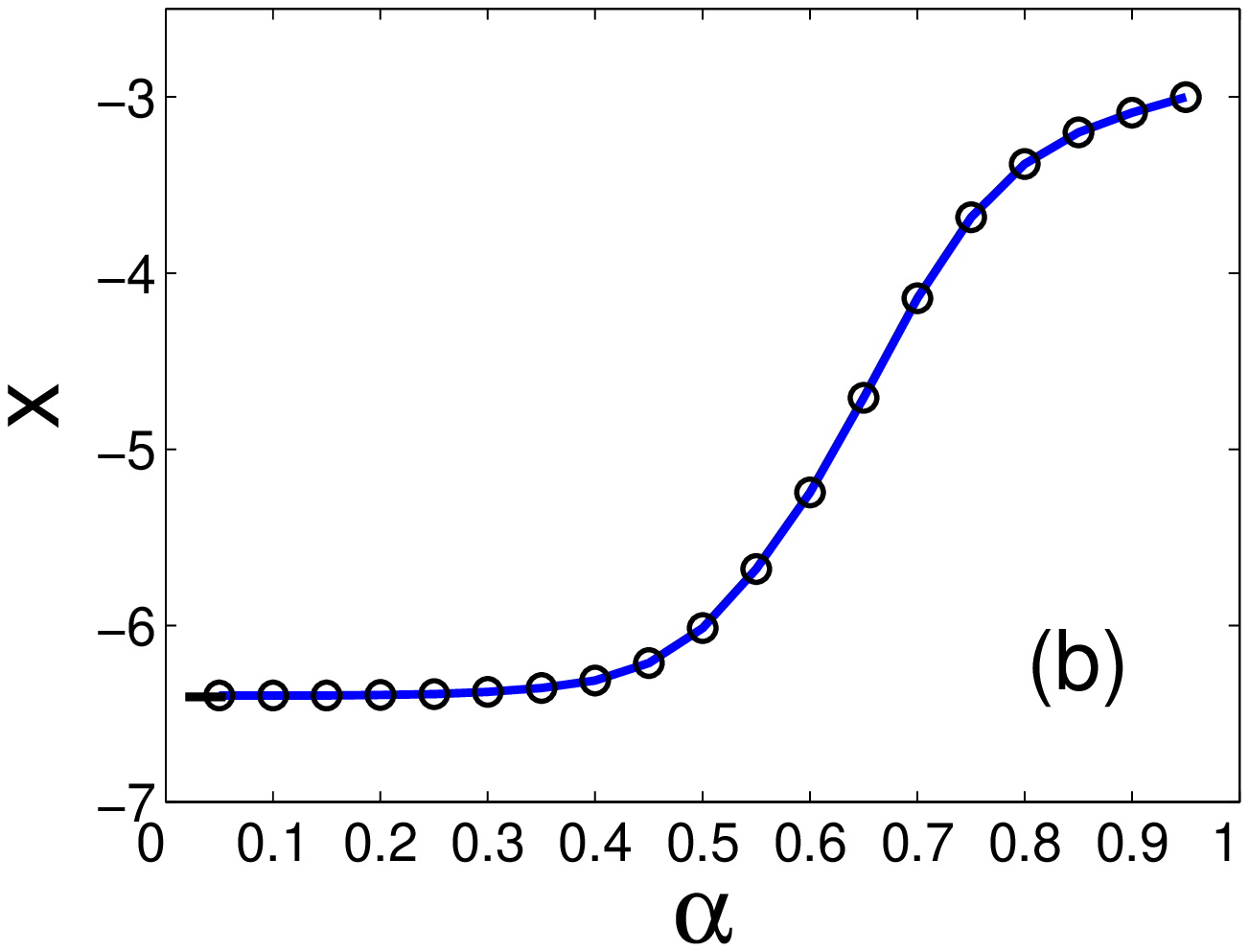}\\
  \caption{Dependence of the output position of solitons on the nonlocality degree
  when the propagation distance is 30. (a) $\delta$=0.1; (b) $\delta$=0.2.
     For both cases,  $p$=1, $\Omega$=0.5, and $\beta_{0}$=0.7.}\label{fig4}
\end{figure}

Fig. 3 shows the dependence of the output position of solitons
 on the nonlocality degree. The curves in Fig. 3 indicate that the nonlocality produces an
important effect on the soliton transverse mobility.  The reason
 is that the Peierls-Nabarro potential barrier for
solitons moving across the lattice is drastically reduced in the
presence of nonlocality, which results in an obvious reduction
of the radiation losses of solitons traveling across the lattices~\cite{Xu-prl-2005}.
Since the nonlocality  makes the radiative
trapping less likely to happen,
an increase of the nonlocality degree leads to an enhancement of the soliton mobility, as shown in Fig. 3.

\section{Conclusions}
In conclusion, we have studied the mobility of solitons in nonlocal nonlinear
 media with a fading optical lattice.
 Based on the Ehrenfest theorem, we obtain an expression of the critical angle for
the soliton's movement in the strongly nonlocal case,
 which decreases with the propagation distance. As a result, the soliton always
 escapes from the central
waveguide as long as the critical angle is smaller than the incident angle.
 It is found through numerical simulations that both the
lattice decay rate and the nonlocality degree have important effects on the soliton
 mobility.

\section*{ACKNOWLEDGEMENT}
The work is supported by
the Doctorial Start-up Fund of Hengyang Normal University, China (Grant No. 11B42), the Natural Science Foundation of Hunan
Province, China (Grant No. 12JJ6001), and
 the construct program of the key discipline in hunan province, China.

\end{document}